\documentclass{article}
\usepackage{amssymb}
\usepackage{amsmath}

 \allowdisplaybreaks

\begin{document}

\title{On the uniqueness of $D=11$ interactions among a graviton, a massless gravitino and a three-form.\\
III: Graviton and gravitini}

\author{E. M. Cioroianu\thanks{ e-mail address:
manache@central.ucv.ro}, E. Diaconu\thanks{ e-mail address:
ediaconu@central.ucv.ro}, S. C. Sararu\thanks{ e-mail
address: scsararu@central.ucv.ro}\\
Faculty of Physics, University of Craiova,\\ 13 Al. I. Cuza Street
Craiova, 200585, Romania}

\maketitle

\begin{abstract}
Under the hypotheses of smoothness of the interactions in the
coupling constant, locality, Poincar\'{e} invariance, Lorentz
covariance, and the preservation of the number of derivatives on
each field in the Lagrangian of the interacting theory (the same
number of derivatives like in the free Lagrangian), we prove that in
$D=11$ there are no cross-interactions between the graviton and the
massless gravitino and also no self-interactions in the
Rarita-Schwinger sector. A comparison with the case $D=4$ is briefly
discussed.


\end{abstract}

\section{Introduction}

Here, we develop the third step of constructing all possible interactions in
$D=11$ among a graviton, a massless Majorana spin-$3/2$ field, and a
three-form gauge field. The previous steps were exposed in~\cite{SUGRAI},
where we obtained all the interactions that can be added to a
eleven-dimensional free theory describing a massless spin-two field and an
Abelian three-form gauge field, and respectively in~\cite{SUGRAII}, where
the same problem was solved with respect to a massless Rarita-Schwinger
field and an Abelian three-form gauge field. Based on the previously
mentioned results, in the sequel we analyze the consistent couplings that
can be introduced between a massless spin-two field (described in the free
limit by the Pauli-Fierz action) and a massless Rarita-Schwinger spinor in
eleven spacetime dimensions. Under the hypotheses of smoothness of the
interactions in the coupling constant, locality, Poincar\'{e} invariance,
Lorentz covariance, and the preservation of the number of derivatives on
each field in the Lagrangian of the interacting theory (the same number of
derivatives like in the free Lagrangian), we prove that in $D=11$ there are
no cross-interactions between the graviton and the massless gravitini and
also no self-interactions among the gravitini. As announced in~\cite{SUGRAII}%
, we comment on the absence of self-interactions among the gravitini in $%
D=11 $ and argue that this result does not contradict the presence in the
Lagrangian of $D=11$, $N=1$ SUGRA of a quartic gravitini vertex. We also
make the comparison with the case $D=4$, where gravitini are known to allow
self-interactions in the presence of a graviton, such that their `mass'
constant becomes related to the cosmological one.

\section{Free model: Lagrangian formulation and BRST symmetry}

Our starting point is represented by a free model, whose lagrangian action
is written as the sum between the action of the linearized version of
Einstein-Hilbert gravity (the Pauli-Fierz action) and that of a massless
Rarita-Schwinger field in eleven spacetime dimensions
\begin{eqnarray}
&&S_{0}^{\mathrm{L}}\left[ h_{\mu \nu },\psi _{\mu }\right] =\int d^{11}x%
\left[ -\frac{1}{2}\left( \partial _{\mu }h_{\nu \rho }\right) \left(
\partial ^{\mu }h^{\nu \rho }\right) +\left( \partial _{\mu }h^{\mu \rho
}\right) \left( \partial ^{\nu }h_{\nu \rho }\right) \right.  \notag \\
&&\left. -\left( \partial _{\mu }h\right) \left( \partial _{\nu }h^{\nu \mu
}\right) +\frac{1}{2}\left( \partial _{\mu }h\right) \left( \partial ^{\mu
}h\right) -\frac{\mathrm{i}}{2}\bar{\psi}_{\mu }\gamma ^{\mu \nu \rho
}\partial _{\nu }\psi _{\rho }\right]  \notag \\
&\equiv &\int d^{11}x\left( \mathcal{L}^{\mathrm{h}}+\mathcal{L}_{0}^{%
\mathrm{\psi }}\right) .  \label{fract}
\end{eqnarray}%
We follow closely all the conventions and notations from~\cite{SUGRAI}
related to the Pauli-Fierz field and respectively from~\cite{SUGRAII} in
relation with the massless Rarita-Schwinger theory. We will need the Fierz
identities specific to $D=11$%
\begin{equation}
\gamma _{\mu _{1}\cdots \mu _{p}}\gamma ^{\nu _{1}\cdots \nu
_{q}}=\sum\limits_{p+q-11\leq 2k\leq 2M}\delta _{\lbrack \mu _{p}}^{[\nu
_{1}}\delta _{\mu _{p-1}}^{\nu _{2}}\cdots \delta _{\mu _{p-k+1}}^{\nu
_{k}}\gamma _{\mu _{1}\cdots \mu _{p-k}]}^{\qquad \qquad \nu _{k+1}\cdots
\nu _{q}]},  \label{fie1}
\end{equation}%
where $M=\min (p,q)$ and also the development of a complex, spinor-like
matrix $N$ in terms of the basis $\left\{ \mathbf{1},\gamma _{\mu },\gamma
_{\mu \nu },\gamma _{\mu \nu \rho },\gamma _{\mu \nu \rho \lambda },\gamma
_{\mu \nu \rho \lambda \sigma }\right\} $%
\begin{equation}
N=\frac{1}{32}\sum\limits_{k=0}^{5}\left( -\right) ^{k(k-1)/2}\frac{1}{k!}%
\mathrm{Tr}\left( \gamma ^{\mu _{1}\cdots \mu _{k}}N\right) \gamma _{\mu
_{1}\cdots \mu _{k}}.  \label{fie2}
\end{equation}%
We recall that $[\mu _{1}\ldots \mu _{k}]$ signifies complete antisymmetry
with respect to the (in this case Lorentz) indices between brackets, with
the conventions that the minimum number of terms is always used and the
result is never divided by the number of terms. Action (\ref{fract})
possesses an irreducible and Abelian generating set of gauge transformations
\begin{equation}
\delta _{\epsilon }h_{\mu \nu }=\partial _{(\mu }\epsilon _{\nu )},\qquad
\delta _{\varepsilon }\psi _{\mu }=\partial _{\mu }\varepsilon ,
\label{pfrs3}
\end{equation}%
with $\epsilon _{\mu }$ bosonic and $\varepsilon $ fermionic gauge
parameters. In addition $\varepsilon $ is a Majorana spinor.

In order to construct the BRST symmetry for (\ref{fract}) we introduce the
fermionic ghosts $\eta _{\mu }$ corresponding to the gauge parameters $%
\epsilon _{\mu }$ and the bosonic, spinor-like ghost $\xi $ corresponding to
the gauge parameter $\varepsilon $ and associate antifields with the
original fields and ghosts, respectively denoted by $\left\{ h^{\ast \mu \nu
},\psi _{\mu }^{\ast }\right\} $ and $\left\{ \eta ^{\ast \mu },\xi ^{\ast
}\right\} $. The antifields of the Rarita-Schwinger fields are bosonic,
purely imaginary spinors. Like in the previous situations (see~\cite%
{SUGRAI,SUGRAII}), the BRST differential simply decomposes into $s=\delta
+\gamma $, where $\delta $ represents the Koszul-Tate differential and $%
\gamma $ stands for the exterior derivative along the gauge orbits. If we
make the notations
\begin{eqnarray}
\Phi ^{A_{0}} &=&\left( h_{\mu \nu },\psi _{\mu }\right) ,\qquad \Phi
_{A_{0}}^{\ast }=\left( h^{\ast \mu \nu },\psi _{\mu }^{\ast }\right) ,
\label{pfrs5} \\
\eta ^{A_{1}} &=&\left( \eta _{\mu },\xi \right) ,\qquad \eta _{A_{1}}^{\ast
}=\left( \eta ^{\ast \mu },\xi ^{\ast }\right) ,  \label{pfrs5a}
\end{eqnarray}%
then, according to the standard rules of the BRST formalism, the degrees of
the BRST generators are valued as: $\mathrm{agh}\left( \Phi ^{A_{0}}\right) =%
\mathrm{agh}\left( \eta ^{A_{1}}\right) =0$, $\mathrm{agh}\left( \Phi
_{A_{0}}^{\ast }\right) =1$, $\mathrm{agh}\left( \eta _{A_{1}}^{\ast
}\right) =2$, $\mathrm{pgh}\left( \Phi ^{A_{0}}\right) =0$, $\mathrm{pgh}%
\left( \eta ^{A_{1}}\right) =1$, $\mathrm{pgh}\left( \Phi _{A_{0}}^{\ast
}\right) =\mathrm{pgh}\left( \eta _{A_{1}}^{\ast }\right) =0$. The actions
of the differentials $\delta $ and $\gamma $ on the generators from the BRST
complex are given by
\begin{eqnarray}
\delta h^{\ast \mu \nu } &=&2H^{\mu \nu },\qquad \delta \psi ^{\ast \mu }=-%
\mathrm{i}\partial _{\rho }\bar{\psi}_{\lambda }\gamma ^{\rho \lambda \mu },
\label{pfrs8} \\
\delta \eta ^{\ast \mu } &=&-2\partial _{\nu }h^{\ast \mu \nu },\qquad
\delta \xi ^{\ast }=\partial _{\mu }\psi ^{\ast \mu },  \label{pfrs9} \\
\delta \Phi ^{A_{0}} &=&0=\delta \eta ^{A_{1}},  \label{pfrs10} \\
\gamma \Phi _{A_{0}}^{\ast } &=&0=\gamma \eta _{A_{1}}^{\ast },
\label{pfrs11} \\
\gamma h_{\mu \nu } &=&\partial _{(\mu }\eta _{\nu )},\qquad \gamma \psi
_{\mu }=\partial _{\mu }\xi ,\qquad \gamma \eta ^{A_{1}}=0,  \label{pfrs12}
\end{eqnarray}%
where $H^{\mu \nu }$ is the linearized Einstein tensor. The full solution to
the master equation $\left( S^{\mathrm{h,\psi }},S^{\mathrm{h,\psi }}\right)
=0$, where $S^{\mathrm{h,\psi }}$ is the anticanonical generator of the BRST
differential with respect to the antibracket structure, $s\cdot =\left(
\cdot ,S^{\mathrm{h,\psi }}\right) $, reads in our case as
\begin{equation}
S^{\mathrm{h,\psi }}=S_{0}^{\mathrm{L}}\left[ h_{\mu \nu },\psi _{\mu }%
\right] +\int d^{11}x\left( h^{\ast \mu \nu }\partial _{(\mu }\eta _{\nu
)}+\psi ^{\ast \mu }\partial _{\mu }\xi \right) .  \label{pfrs16}
\end{equation}

\section{Consistent interactions between a graviton and gravitini}

In order to investigate the consistent couplings that can be added to the
free action (\ref{fract}) we act like in~\cite{SUGRAI,SUGRAII} and rely on
the reformulation of the interaction problem in the context of the
antifield-BRST deformation procedure. Thus, if an interacting gauge theory
can be consistently constructed, then the solution $S^{\mathrm{h,\psi }}$ to
the master equation associated with the free theory, (\ref{pfrs16}), can be
deformed into a solution $\bar{S}^{\mathrm{h,\psi }}$
\begin{eqnarray}
S^{\mathrm{h,\psi }} &\rightarrow &\bar{S}^{\mathrm{h,\psi }}=S^{\mathrm{%
h,\psi }}+\lambda S_{1}^{\mathrm{h,\psi }}+\lambda ^{2}S_{2}^{\mathrm{h,\psi
}}+\cdots  \notag \\
&=&S^{\mathrm{h,\psi }}+\lambda \int d^{D}x\,a^{\mathrm{h,\psi }}+\lambda
^{2}\int d^{D}x\,b^{\mathrm{h,\psi }}+\cdots  \label{2.2}
\end{eqnarray}%
of the master equation for the deformed theory $\left( \bar{S}^{\mathrm{%
h,\psi }},\bar{S}^{\mathrm{h,\psi }}\right) =0$, such that both the ghost
and antifield spectra of the initial theory are preserved. The last equation
splits, according to the various orders in the coupling constant $\lambda $,
into the equivalent tower of equations: $\left( S^{\mathrm{h,\psi }},S^{%
\mathrm{h,\psi }}\right) =0$ and%
\begin{eqnarray}
2\left( S_{1}^{\mathrm{h,\psi }},S^{\mathrm{h,\psi }}\right) &=&0,
\label{2.5} \\
2\left( S_{2}^{\mathrm{h,\psi }},S^{\mathrm{h,\psi }}\right) +\left( S_{1}^{%
\mathrm{h,\psi }},S_{1}^{\mathrm{h,\psi }}\right) &=&0,  \label{2.6} \\
&&\vdots  \notag
\end{eqnarray}%
Equation $\left( S^{\mathrm{h,\psi }},S^{\mathrm{h,\psi }}\right) =0$ is
fulfilled by hypothesis, while the next one requires that the first-order
deformation of the solution to the master equation, $S_{1}^{\mathrm{h,\psi }%
} $, is a (nontrivial) co-cycle of the \textquotedblleft
free\textquotedblright\ BRST differential at ghost number zero, $S_{1}^{%
\mathrm{h,\psi }}\in H^{0}\left( s\right) $. Our main concern is to
determine $S_{1}^{\mathrm{h,\psi }}$, $S_{2}^{\mathrm{h,\psi }}$, etc. that
comply with all the main hypotheses: smoothness of interactions in the
coupling constant, locality, Poincar\'{e} invariance, Lorentz covariance,
and the preservation of the number of derivatives on each field in the
interacting Lagrangian with respect to the free theory.

\section{First-order deformation}

The resolution of equation (\ref{2.5}) implies standard cohomological
techniques related to the BRST differential of the free model under
consideration. The necessary cohomological ingredients have already been
discussed in~\cite{SUGRAI,SUGRAII}, so in the sequel we give the solutions
to these equations without going into further details. The nontrivial
solution to equation (\ref{2.5}) can be shown to expand as $S_{1}^{\mathrm{%
h,\psi }}=\int d^{11}x\left( a_{0}^{\mathrm{h,\psi }}+a_{1}^{\mathrm{h,\psi }%
}+a_{2}^{\mathrm{h,\psi }}\right) $, where $\mathrm{agh}\left( a_{k}^{%
\mathrm{h,\psi }}\right) =k$. We can further decompose $a_{k}^{\mathrm{%
h,\psi }}$ in a natural manner as a sum between three kinds of deformations
\begin{equation}
a_{k}^{\mathrm{h,\psi }}=a_{k}^{\mathrm{h}}+a_{k}^{\mathrm{h-\psi }}+a_{k}^{%
\mathrm{\psi }},  \label{pfrs3.12a}
\end{equation}%
where $a_{k}^{\mathrm{h}}$ contains only fields/ghosts/antifields from the
Pauli-Fierz sector, $a_{k}^{\mathrm{h-\psi }}$ describes the
cross-interactions between the two theories (so it effectively mixes both
sectors), and $a_{k}^{\mathrm{\psi }}$ involves only the Rarita-Schwinger
sector. The components $a_{2}^{\mathrm{h}}$ and $a_{1}^{\mathrm{h}}$ are
given by
\begin{eqnarray}
a_{2}^{\mathrm{h}} &=&\frac{1}{2}\eta ^{\ast \mu }\eta ^{\nu }\partial _{%
\left[ \mu \right. }\eta _{\left. \nu \right] },  \label{PFa2} \\
a_{1}^{\mathrm{h}} &=&h^{\ast \mu \rho }\left( \left( \partial _{\rho }\eta
^{\nu }\right) h_{\mu \nu }-\eta ^{\nu }\partial _{\lbrack \mu }h_{\nu ]\rho
}\right) ,  \label{PFa1}
\end{eqnarray}%
while $a_{0}^{\mathrm{h}}$ is the cubic vertex of the Einstein-Hilbert
lagrangian plus a cosmological term
\begin{equation}
a_{0}^{\mathrm{h}}=a_{0}^{\mathrm{h-cubic}}-2\Lambda h,  \label{PFa0}
\end{equation}%
with $\Lambda $ the cosmological constant. Related to the interaction
sector, it can be shown that the components $a_{k}^{\mathrm{h-\psi }}$read
as
\begin{equation}
a_{2}^{\mathrm{h-\psi }}=\frac{\bar{k}}{8}\left( -\frac{\mathrm{i}}{2}\eta
^{\ast \mu }\bar{\xi}\gamma _{\mu }\xi +\xi ^{\ast }\gamma ^{\mu \nu }\xi
\partial _{\lbrack \mu }\eta _{\nu ]}\right) ,  \label{pfrs3.33a}
\end{equation}%
\begin{eqnarray}
a_{1}^{\mathrm{h-\psi }} &=&\frac{\bar{k}}{4}\left[ \mathrm{i}h^{\ast \mu
\nu }\bar{\xi}\gamma _{\mu }\psi _{\nu }+\frac{1}{2}\psi ^{\ast \mu }\gamma
^{\alpha \beta }\left( \psi _{\mu }\partial _{\lbrack \alpha }\eta _{\beta
]}-\xi \partial _{\lbrack \alpha }h_{\beta ]\mu }\right) \right.  \notag \\
&&\left. -4\psi ^{\ast \mu }\left( \partial _{\lbrack \mu }\psi _{\nu
]}\right) \eta ^{\nu }\right] ,  \label{pfrs3.33b}
\end{eqnarray}%
\begin{eqnarray}
a_{0}^{\mathrm{h-\psi }} &=&\frac{\mathrm{i}\bar{k}}{4}\left[ \left(
\partial _{\alpha }\bar{\psi}_{\beta }\right) \gamma ^{\alpha \beta \mu
}\psi ^{\nu }\left( h_{\mu \nu }-\sigma _{\mu \nu }h\right) +\left( \partial
_{\lbrack \alpha }\bar{\psi}_{\mu ]}\right) \gamma ^{\mu \nu \beta }\psi
_{\nu }h_{\ \ \beta }^{\alpha }\right.  \notag \\
&&\left. +\frac{1}{2}\bar{\psi}^{\mu }\left( \gamma ^{\rho }\psi ^{\nu
}-2\sigma ^{\nu \rho }\gamma _{\lambda }\psi ^{\lambda }\right) \partial
_{\lbrack \mu }h_{\nu ]\rho }\right] ,  \label{pfrs3.33c}
\end{eqnarray}%
while for the self-interactions of the Rarita-Schwinger field one obtains
\begin{eqnarray}
a_{2}^{\mathrm{\psi }} &=&0,\qquad a_{1}^{\mathrm{\psi }}=\mathrm{i}m\psi
_{\mu }^{\ast }\gamma ^{\mu }\xi ,  \label{qw1} \\
a_{0}^{\mathrm{\psi }} &=&-\frac{9m}{2}\bar{\psi}_{\mu }\gamma ^{\mu \nu
}\psi _{\nu },  \label{qw2}
\end{eqnarray}%
with $\bar{k}$ and $m$ some arbitrary, real constants.

\section{Second-order deformation}

We have seen in the above that the first-order deformation can be written as
the sum between the first-order deformation of the solution to the master
equation for the Pauli-Fierz theory $S_{1}^{\mathrm{h}}=\int d^{11}x\left(
a_{2}^{\mathrm{h}}+a_{1}^{\mathrm{h}}+a_{0}^{\mathrm{h}}\right) $, the
`interacting' part $S_{1}^{\mathrm{h-\psi }}=\int d^{11}x\left( a_{2}^{%
\mathrm{h-\psi }}+a_{1}^{\mathrm{h-\psi }}+a_{0}^{\mathrm{h-\psi }}\right) $%
, and the Rarita-Schwinger component $S_{1}^{\mathrm{\psi }}=\int
d^{11}x\left( a_{1}^{\mathrm{\psi }}+a_{0}^{\mathrm{\psi }}\right) $. Thus,
the first-order deformation is parameterized in terms of three real
constants: $\Lambda $, $\bar{k}$, and $m$.

In the sequel we infer the complete expression of the second-order
deformation of the solution to the master equation, $S_{2}^{\mathrm{h,\psi }%
} $, which is subject to equation (\ref{2.6}). Acting like in the above, we
can write the second-order deformation as the sum between the Pauli-Fierz,
the Rarita-Schwinger, and the interacting parts, $S_{2}^{\mathrm{h,\psi }%
}=S_{2}^{\mathrm{h}}+\bar{S}_{2}^{\mathrm{\psi }}+S_{2}^{\mathrm{h-\psi }}$.
The piece $S_{2}^{\mathrm{h}}$ describes the second-order deformation in the
Pauli-Fierz sector and we will not insist on it since we are merely
interested in cross-couplings. The term $\bar{S}_{2}^{\mathrm{\psi }}$
results as solution to the equation
\begin{equation}
\left( S_{1},S_{1}\right) ^{\mathrm{\psi }}+2s\bar{S}_{2}^{\mathrm{\psi }}=0,
\label{pfrs4.4}
\end{equation}%
where $\left( S_{1},S_{1}\right) ^{\mathrm{\psi }}=\left( S_{1}^{\mathrm{%
\psi }},S_{1}^{\mathrm{\psi }}\right) +\left( S_{1}^{\mathrm{h-\psi }%
},S_{1}^{\mathrm{h-\psi }}\right) ^{\mathrm{\psi }}$. In the last formula
the notation $\left( S_{1}^{\mathrm{h-\psi }},S_{1}^{\mathrm{h-\psi }%
}\right) ^{\mathrm{\psi }}$ signifies the terms from the antibracket $\left(
S_{1}^{\mathrm{h-\psi }},S_{1}^{\mathrm{h-\psi }}\right) $ that contain only
BRST generators from the Rarita-Schwinger sector. The piece $S_{2}^{\mathrm{%
h-\psi }}$ is solution to the equation
\begin{equation}
\left( S_{1},S_{1}\right) ^{\mathrm{h-\psi }}+2sS_{2}^{\mathrm{h-\psi }}=0,
\label{pfrs4.6}
\end{equation}%
where $\left( S_{1},S_{1}\right) ^{\mathrm{h-\psi }}=2\left( S_{1}^{\mathrm{%
\psi }},S_{1}^{\mathrm{h-\psi }}\right) +2\left( S_{1}^{\mathrm{h}},S_{1}^{%
\mathrm{h-\psi }}\right) +\left( S_{1}^{\mathrm{h-\psi }},S_{1}^{\mathrm{%
h-\psi }}\right) ^{\mathrm{h-\psi }}$. In the last relation we used the
notation $\left( S_{1}^{\mathrm{h-\psi }},S_{1}^{\mathrm{h-\psi }}\right) ^{%
\mathrm{h-\psi }}=\left( S_{1}^{\mathrm{h-\psi }},S_{1}^{\mathrm{h-\psi }%
}\right) -\left( S_{1}^{\mathrm{h-\psi }},S_{1}^{\mathrm{h-\psi }}\right) ^{%
\mathrm{\psi }}$. If we denote by $\bar{\Delta}^{\mathrm{\psi }}$ and $\bar{b%
}^{\mathrm{\psi }}$ the nonintegrated densities of $\left(
S_{1},S_{1}\right) ^{\mathrm{\psi }}$ and respectively of $\bar{S}_{2}^{%
\mathrm{\psi }}$, then equation (\ref{pfrs4.4}) takes the local form
\begin{equation}
\bar{\Delta}^{\mathrm{\psi }}=-2s\bar{b}^{\mathrm{\psi }}+\partial _{\mu }%
\bar{n}^{\mu },  \label{pfrs4.8}
\end{equation}%
with $\mathrm{gh}\left( \bar{\Delta}^{\mathrm{\psi }}\right) =1$, $\mathrm{gh%
}\left( \bar{b}^{\mathrm{\psi }}\right) =0$, $\mathrm{gh}\left( \bar{n}^{\mu
}\right) =1$, for some local currents $\bar{n}^{\mu }$. Direct computation
shows that $\bar{\Delta}^{\mathrm{\psi }}$ decomposes as $\bar{\Delta}^{%
\mathrm{\psi }}=\bar{\Delta}_{0}^{\mathrm{\psi }}+\bar{\Delta}_{1}^{\mathrm{%
\psi }}+\bar{\Delta}_{2}^{\mathrm{\psi }}$, with $\mathrm{agh}\left( \bar{%
\Delta}_{I}^{\mathrm{\psi }}\right) =I$, $I=\overline{0,2}$, where
\begin{equation}
\bar{\Delta}_{2}^{\mathrm{\psi }}=\partial _{\mu }\bar{\tau}_{2}^{\mu
}+\gamma \left( \frac{\mathrm{i}\bar{k}^{2}}{16}\xi ^{\ast }\gamma ^{\mu \nu
}\xi \bar{\xi}\gamma _{\mu }\psi _{\nu }\right) ,  \label{pfrs4.11}
\end{equation}%
\begin{eqnarray}
\bar{\Delta}_{1}^{\mathrm{\psi }} &=&\partial _{\mu }\bar{\tau}_{1}^{\mu
}+\delta \left( \frac{\mathrm{i}\bar{k}^{2}}{16}\xi ^{\ast }\gamma ^{\mu \nu
}\xi \bar{\xi}\gamma _{\mu }\psi _{\nu }\right) +\gamma \left[ \frac{\mathrm{%
i}\bar{k}^{2}}{16}\psi ^{\ast \mu }\gamma ^{\alpha \beta }\psi _{\mu }\bar{%
\xi}\gamma _{\alpha }\psi _{\beta }\right.  \notag \\
&&\left. -\frac{\mathrm{i}\bar{k}^{2}}{16}\psi ^{\ast \mu }\gamma ^{\alpha
\beta }\xi \left( \bar{\psi}_{\mu }\gamma _{\alpha }\psi _{\beta }+\frac{1}{2%
}\bar{\psi}_{\alpha }\gamma _{\mu }\psi _{\beta }\right) \right]  \notag \\
&&+\frac{\mathrm{i}\bar{k}^{2}}{8}\psi ^{\ast \mu }\left[ \left( \partial
_{\lbrack \mu }\psi _{\nu ]}\right) \bar{\xi}\gamma ^{\nu }\xi \right.
\notag \\
&&\left. +\frac{1}{2}\left( \gamma ^{\nu \rho }\xi \right) \left( \bar{\xi}%
\gamma _{\rho }\partial _{\lbrack \mu }\psi _{\nu ]}-\frac{1}{2}\bar{\xi}%
\gamma _{\mu }\partial _{\lbrack \nu }\psi _{\rho ]}\right) \right] ,
\label{pfrs4.12}
\end{eqnarray}%
\begin{eqnarray}
\bar{\Delta}_{0}^{\mathrm{\psi }} &=&\partial _{\mu }\bar{\tau}_{0}^{\mu }+%
\mathrm{i}\left( 180m^{2}-\bar{k}\Lambda \right) \bar{\xi}\gamma ^{\mu }\psi
_{\mu }  \notag \\
&&-\frac{\bar{k}^{2}}{16}\left[ \left( \partial _{\alpha }\bar{\psi}_{\beta
}\right) \gamma ^{\alpha \beta \mu }\psi ^{\nu }\left( \bar{\xi}\gamma
_{(\mu }\psi _{\nu )}-2\sigma _{\mu \nu }\bar{\xi}\gamma ^{\rho }\psi _{\rho
}\right) \right.  \notag \\
&&+\bar{\psi}_{\mu }\gamma ^{\mu \nu \rho }\left( \partial _{\lbrack \nu
}\psi _{\alpha ]}\right) \bar{\xi}\gamma ^{(\alpha }\psi ^{\beta )}\sigma
_{\rho \beta }+\bar{\psi}^{\alpha }\gamma ^{\rho }\psi _{\rho }\partial
^{\beta }\left( \bar{\xi}\gamma _{(\alpha }\psi _{\beta )}\right)  \notag \\
&&\left. +\bar{\psi}^{\alpha }\gamma ^{\rho }\psi ^{\beta }\partial _{\alpha
}\left( \bar{\xi}\gamma _{(\beta }\psi _{\rho )}-2\sigma _{\rho \beta }\bar{%
\xi}\gamma ^{\lambda }\psi _{\lambda }\right) \right] .  \label{pfrs4.13}
\end{eqnarray}

Since $\bar{\Delta}^{\mathrm{\psi }}$ stops at antighost number two, we can
take, without loss of generality, the corresponding second-order deformation
to stop at antighost number three, $\bar{b}^{\mathrm{\psi }%
}=\sum\limits_{I=0}^{3}\bar{b}_{I}^{\mathrm{\psi }}$, $\mathrm{agh}\left(
\bar{b}_{I}^{\mathrm{\psi }}\right) =I$, $I=\overline{0,3}$, $\bar{n}^{\mu
}=\sum\limits_{I=0}^{3}\bar{n}_{I}^{\mu }$, $\mathrm{agh}\left( \bar{n}%
_{I}^{\mu }\right) =I$, $I=\overline{0,3}$. By projecting (\ref{pfrs4.8}) on
the various (decreasing) values of the antighost number, we obtain the
equivalent tower of equations
\begin{eqnarray}
0 &=&-2\gamma \bar{b}_{3}^{\mathrm{\psi }}+\partial _{\mu }\bar{n}_{3}^{\mu
},  \label{pfrs4.16a} \\
\bar{\Delta}_{2}^{\mathrm{\psi }} &=&-2\left( \delta \bar{b}_{3}^{\mathrm{%
\psi }}+\gamma \bar{b}_{2}^{\mathrm{\psi }}\right) +\partial _{\mu }\bar{n}%
_{2}^{\mu },  \label{pfrs4.16b} \\
\bar{\Delta}_{1}^{\mathrm{\psi }} &=&-2\left( \delta \bar{b}_{2}^{\mathrm{%
\psi }}+\gamma \bar{b}_{1}^{\mathrm{\psi }}\right) +\partial _{\mu }\bar{n}%
_{1}^{\mu },  \label{pfrs4.16c} \\
\bar{\Delta}_{0}^{\mathrm{\psi }} &=&-2\left( \delta \bar{b}_{1}^{\mathrm{%
\psi }}+\gamma \bar{b}_{0}^{\mathrm{\psi }}\right) +\partial _{\mu }\bar{n}%
_{0}^{\mu }.  \label{pfrs4.16d}
\end{eqnarray}
Equation (\ref{pfrs4.16a}) can always be replaced, by adding trivial terms
only, with $\gamma \bar{b}_{3}^{\mathrm{\psi }}=0$. Thus, $\bar{b}_{3}^{%
\mathrm{\psi }}$ belongs to the Rarita-Schwinger sector of cohomology of $%
\gamma $, $H\left( \gamma \right) $. By means of definitions (\ref{pfrs11}%
)--(\ref{pfrs12}) we get that $H\left( \gamma \right) $ in the
Rarita-Schwinger sector is generated by the objects $\left( \psi ^{*\mu
},\xi ^{*},\partial _{[\mu }\psi _{\nu ]}\right) $, by their spacetime
derivatives up to a finite order, and also by the undifferentiated ghosts $%
\xi $ (the spacetime derivatives of $\xi $ are $\gamma $-exact according to
the second relation in (\ref{pfrs3})). As a consequence, we can write $\bar{b%
}_{3}^{\mathrm{\psi }}=\bar{\beta}_{3}^{\mathrm{\psi }}\left( \left[
\partial _{[\mu }\psi _{\nu ]}\right] ,\left[ \psi ^{*\mu }\right] ,\left[
\xi ^{*}\right] \right) e^{3}\left( \xi \right) $, where $e^{3}\left( \xi
\right) $ are the elements of pure ghost number three of a basis in the
space of polynomials in the ghosts $\xi $ and the notation $f\left( \left[ q%
\right] \right) $ means that $f$ depends on $q$ and its spacetime
derivatives up to a finite order. Inserting (\ref{pfrs4.11}) in (\ref%
{pfrs4.16b}) and using standard cohomological arguments, we reach the
conclusion that $\bar{\beta}_{3}^{\mathrm{\psi }}$ are (nontrivial) elements
of $H_{3}^{\mathrm{inv}}\left( \delta |d\right) $, where $H_{3}^{\mathrm{inv}%
}\left( \delta |d\right) $ denotes as usually the local cohomology
of the Koszul-Tate differential in the space of invariant
polynomials in antighost number three for the free theory
(\ref{fract}). (By `invariant polynomials' we mean elements of
$H\left( \gamma \right) $ at pure ghost number zero.) On the other
hand, $H_{3}^{\mathrm{inv}}\left( \delta |d\right) =0$ for the
free theory under consideration, such that we can safely take $\bar{\beta}%
_{3}^{\mathrm{\psi }}=0$, which further leads to $\bar{b}_{3}^{\mathrm{\psi }%
}=0$.

With this result at hand, from (\ref{pfrs4.16b}) and (\ref{pfrs4.11}) it
follows that
\begin{equation}
\bar{b}_{2}^{\mathrm{\psi }}=-\frac{\mathrm{i}\bar{k}^{2}}{32}\xi ^{\ast
}\gamma ^{\mu \nu }\xi \bar{\xi}\gamma _{\mu }\psi _{\nu }+\tilde{b}_{2}^{%
\mathrm{\psi }},  \label{pfrs4.19}
\end{equation}%
where $\tilde{b}_{2}^{\mathrm{\psi }}$ is solution to the equation $\gamma
\tilde{b}_{2}^{\mathrm{\psi }}=0$. Looking at $\bar{\Delta}_{1}^{\mathrm{%
\psi }}$ given in (\ref{pfrs4.12}), it results that it can be written as in (%
\ref{pfrs4.16c}) if
\begin{equation}
\bar{\chi}=\frac{\mathrm{i}\bar{k}^{2}}{8}\psi ^{\ast \mu }\left[ \left(
\partial _{\lbrack \mu }\psi _{\nu ]}\right) \bar{\xi}\gamma ^{\nu }\xi +%
\frac{1}{2}\left( \gamma ^{\nu \rho }\xi \right) \left( \bar{\xi}\gamma
_{\rho }\partial _{\lbrack \mu }\psi _{\nu ]}-\frac{1}{2}\bar{\xi}\gamma
_{\mu }\partial _{\lbrack \nu }\psi _{\rho ]}\right) \right] ,
\label{pfrs4.20}
\end{equation}%
can be expressed like
\begin{equation}
\bar{\chi}=-2\delta \tilde{b}_{2}^{\mathrm{\psi }}+\gamma \bar{\rho}%
+\partial _{\mu }\bar{l}^{\mu },  \label{pfrs4.21}
\end{equation}%
where
\begin{equation}
\bar{\rho}=-\frac{\mathrm{i}\bar{k}^{2}}{16}\psi ^{\ast \mu }\gamma ^{\alpha
\beta }\left( \psi _{\mu }\bar{\xi}\gamma _{\alpha }\psi _{\beta }-\xi
\left( \bar{\psi}_{\mu }\gamma _{\alpha }\psi _{\beta }+\frac{1}{2}\bar{\psi}%
_{\alpha }\gamma _{\mu }\psi _{\beta }\right) \right) -2\bar{b}_{1}^{\mathrm{%
\psi }}.  \label{mn1}
\end{equation}%
Assume that (\ref{pfrs4.21}) holds. Then, by taking its left Euler-Lagrange
(EL) derivatives with respect to $\psi ^{\ast \mu }$ and using the
commutation between $\gamma $ and each EL derivative $\delta ^{\mathrm{L}%
}/\delta \psi _{\mu }^{\ast }$, we infer the relations
\begin{equation}
\frac{\delta ^{\mathrm{L}}\left( \bar{\chi}+2\delta \tilde{b}_{2}^{\mathrm{%
\psi }}\right) }{\delta \psi ^{\ast \mu }}=\gamma \left( \frac{\delta ^{%
\mathrm{L}}\bar{\rho}}{\delta \psi ^{\ast \mu }}\right) .  \label{pfrs4.21e}
\end{equation}%
As $\tilde{b}_{2}^{\mathrm{\psi }}$ is $\gamma $-invariant, then $\delta
\tilde{b}_{2}^{\mathrm{\psi }}$ will also be $\gamma $-invariant. Recalling
the previous results on the cohomology of $\gamma $ in the Rarita-Schwinger
sector, we find that $\delta \tilde{b}_{2}^{\mathrm{\psi }}=e^{2}\left( \xi
\right) \psi ^{\ast \mu }v_{\mu }$, with $v_{\mu }$ fermionic, $\gamma $%
-invariant functions of antighost number zero and $e^{2}\left( \xi \right) $
the elements of pure ghost number two of a basis in the space of polynomials
in the ghosts $\xi $ . By using (\ref{pfrs4.20}) and the last expression of $%
\delta \tilde{b}_{2}^{\mathrm{\psi }}$, direct computation provides the
equation
\begin{eqnarray}
&&\frac{\delta ^{\mathrm{L}}\left( \bar{\chi}+2\delta \tilde{b}_{2}^{\mathrm{%
\psi }}\right) }{\delta \psi ^{\ast \mu }}=\frac{\mathrm{i}\bar{k}^{2}}{8}%
\left[ \frac{1}{2}\left( \gamma ^{\nu \rho }\xi \right) \left( \bar{\xi}%
\gamma _{\rho }\partial _{\lbrack \mu }\psi _{\nu ]}-\frac{1}{2}\bar{\xi}%
\gamma _{\mu }\partial _{\lbrack \nu }\psi _{\rho ]}\right) \right.  \notag
\\
&&\left. +\left( \partial _{\lbrack \mu }\psi _{\nu ]}\right) \bar{\xi}%
\gamma ^{\nu }\xi \right] +2e^{2}\left( \xi \right) v_{\mu }.  \label{x}
\end{eqnarray}%
On the one hand, equation (\ref{pfrs4.21e}) shows that $\delta ^{\mathrm{L}%
}\left( \bar{\chi}+2\delta \tilde{b}_{2}^{\mathrm{\psi }}\right) /\delta
\psi ^{\ast \mu }$ is trivial in $H\left( \gamma \right) $. On the other
hand, relation (\ref{x}) emphasizes that $\delta ^{\mathrm{L}}\left( \bar{%
\chi}+2\delta \tilde{b}_{2}^{\mathrm{\psi }}\right) /\delta \psi ^{\ast \mu
} $ is a nontrivial element from $H\left( \gamma \right) $ (because each
term in the right-hand side of (\ref{x}) is nontrivial in $H\left( \gamma
\right) $). Then, $\delta ^{\mathrm{L}}\left( \bar{\chi}+2\delta \tilde{b}%
_{2}^{\mathrm{\psi }}\right) /\delta \psi ^{\ast \mu }$ must be set zero
\begin{equation}
\frac{\delta ^{\mathrm{L}}\left( \bar{\chi}+2\delta \tilde{b}_{2}^{\mathrm{%
\psi }}\right) }{\delta \psi ^{\ast \mu }}=0,  \label{mn2}
\end{equation}%
which yields\footnote{%
In fact, the general solution to equation (\ref{mn3}) takes the form $\bar{%
\chi}+2\delta \tilde{b}_{2}^{\mathrm{\psi}}=u+\partial _{\mu }\tilde{l}^{\mu }$%
, where $u$ is a function of antighost number one depending on all the BRST
generators from the Rarita-Schwinger sector but the antifields $\psi ^{*\mu
} $. As the antifields $\psi ^{*\mu }$ are the only Rarita-Schwinger
antifields of antighost number one, the condition $\mathrm{agh}\left(
u\right) =1$ automatically produces $u=0$.}
\begin{equation}
\bar{\chi}+2\delta \tilde{b}_{2}^{\mathrm{\psi }}=\partial _{\mu }\tilde{l}%
^{\mu }.  \label{mn3}
\end{equation}%
By acting with $\delta $ on (\ref{mn3}) we deduce
\begin{equation}
\delta \bar{\chi}=\partial _{\mu }\bar{j}^{\mu }.  \label{mn4}
\end{equation}%
From (\ref{pfrs4.20}), by direct computation we find
\begin{equation}
\delta \bar{\chi}=-\frac{\bar{k}^{2}}{8}\left( \partial _{\alpha }\bar{\psi}%
_{\lambda }\right) \gamma ^{\alpha \lambda \mu }\left[ \left( \partial
_{\lbrack \mu }\psi _{\nu ]}\right) \bar{\xi}\gamma ^{\nu }\xi +\frac{1}{2}%
\left( \gamma ^{\nu \rho }\xi \right) \left( \bar{\xi}\gamma _{\rho
}\partial _{\lbrack \mu }\psi _{\nu ]}-\frac{1}{2}\bar{\xi}\gamma _{\mu
}\partial _{\lbrack \nu }\psi _{\rho ]}\right) \right] .  \label{mn5}
\end{equation}%
Comparing (\ref{mn4}) with (\ref{mn5}) and recalling the Noether identities
corresponding to the Rarita-Schwinger action, we obtain that the right-hand
of (\ref{mn5}) reduces to a total derivative iff
\begin{equation}
\left( \partial _{\lbrack \mu }\psi _{\nu ]}\right) \bar{\xi}\gamma ^{\nu
}\xi +\frac{1}{2}\left( \gamma ^{\nu \rho }\xi \right) \left( \bar{\xi}%
\gamma _{\rho }\partial _{\lbrack \mu }\psi _{\nu ]}-\frac{1}{2}\bar{\xi}%
\gamma _{\mu }\partial _{\lbrack \nu }\psi _{\rho ]}\right) =\partial _{\mu }%
\bar{p}.  \label{mn6}
\end{equation}%
Simple computation exhibits that the left-hand side of (\ref{mn6}) cannot be
written like a total derivative, so neither relation (\ref{mn4}) nor
equation (\ref{pfrs4.21}) hold. As a consequence, $\bar{\chi}$ must vanish
and hence we must set
\begin{equation}
\bar{k}=0.  \label{pfrs4.22}
\end{equation}

Inserting (\ref{pfrs4.22}) in (\ref{pfrs4.11})--(\ref{pfrs4.13}), we obtain
that
\begin{eqnarray}
\bar{\Delta}_{2}^{\mathrm{\psi }} &=&\partial _{\mu }\bar{\tau}_{2}^{\mu
},\qquad \bar{\Delta}_{1}^{\mathrm{\psi }}=\partial _{\mu }\bar{\tau}%
_{1}^{\mu },  \label{pfrs4.33a} \\
\bar{\Delta}_{0}^{\mathrm{\psi }} &=&\partial _{\mu }\bar{\tau}_{0}^{\mu
}+180\mathrm{i}m^{2}\bar{\xi}\gamma ^{\mu }\psi _{\mu }.  \label{pfrs4.33b}
\end{eqnarray}%
From (\ref{pfrs4.33a}) it results that we can safely take $\bar{b}_{2}^{%
\mathrm{\psi }}=0$ and $\bar{b}_{1}^{\mathrm{\psi }}=0$, which replaced in (%
\ref{pfrs4.16d}) lead to the necessary condition that $\bar{\Delta}_{0}^{%
\mathrm{\psi }}$ must be a trivial element from the local cohomology of $%
\gamma $, i.e. $\bar{\Delta}_{0}^{\mathrm{\psi }}=-2\gamma \bar{b}_{0}^{%
\mathrm{\psi }}+\partial _{\mu }\bar{n}_{0}^{\mu }$. In order to solve this
equation with respect to $\bar{b}_{0}^{\mathrm{\psi }}$, we will project it
on the number of derivatives. Since $\gamma \bar{b}_{0}^{\mathrm{\psi }}$
contains at least one spacetime derivative, the above equation projected on
the number of derivatives equal to zero reduces to $\bar{\Delta}_{0}^{%
\mathrm{\psi }}=180\mathrm{i}m^{2}\bar{\xi}\gamma ^{\mu }\psi _{\mu }=0$,
which further implies
\begin{equation}
m=0.  \label{wzz1}
\end{equation}%
Substituting relations (\ref{pfrs4.22}) and (\ref{wzz1}) in (\ref{pfrs3.33a}%
)--(\ref{qw2}) we obtain that $S_{1}^{\mathrm{h-\psi }}=0$ and $S_{1}^{%
\mathrm{\psi }}=0$, so equations (\ref{pfrs4.4})--(\ref{pfrs4.6}) possess
only the trivial solution $S_{2}^{\mathrm{h-\psi }}=0$ and $S_{2}^{\mathrm{%
\psi }}=0$. The vanishing of $S_{1}^{\mathrm{h-\psi }}$, $S_{2}^{\mathrm{%
h-\psi }}$, $S_{1}^{\mathrm{\psi }}$, and $S_{2}^{\mathrm{\psi }}$ further
leads, via the equations that stipulate the higher-order deformation
equations, to the result that we can take
\begin{equation}
S_{i}^{\mathrm{h-\psi }}=0,\qquad S_{i}^{\mathrm{\psi }}=0,\qquad i\geq 1.
\label{r160c}
\end{equation}%
In conclusion, under the hypotheses of locality, smoothness of the
interactions in the coupling constant, Poincar\'{e} invariance, (background)
Lorentz invariance, and the preservation of the number of derivatives on
each field, there are no cross-interactions between the Pauli-Fierz field
and the massless Rarita-Schwinger field and also no self-interactions for
the massless Rarita-Schwinger field, both in $D=11$.

\section{On the quartic SUGRA gravitini vertex}

We have seen (here and in \cite{SUGRAII}) that gravitini allows no
self-interactions in $D=11$ if separately coupled to a graviton or
respectively to a three-form gauge field. Nevertheless, it is known that $%
D=11$, $N=1$ SUGRA contains a quartic vertex (at order two in the coupling
constant $\lambda $) expressing self-interactions among the gravitini. This
apparent paradox has a simple explanation. If we start from a free theory
with \textit{all} the three fields (gravitini, graviton, and three-form),
then the consistency of the first-order deformation in the Rarita-Schwinger
sector becomes
\begin{equation}
\left( S_{1},S_{1}\right) ^{\mathrm{\psi }}+2sS_{2}^{\mathrm{\psi }}=0,
\label{all1}
\end{equation}%
where $\left( S_{1},S_{1}\right) ^{\mathrm{\psi }}$ collects now all the
self-interaction terms, coming from $\left( S_{1}^{\mathrm{\psi }},S_{1}^{%
\mathrm{\psi }}\right) $, $\left( S_{1}^{\mathrm{h-\psi }},S_{1}^{\mathrm{%
h-\psi }}\right) $, \textit{and} $\left( S_{1}^{\mathrm{A-\psi }},S_{1}^{%
\mathrm{A-\psi }}\right) $ (where $S_{1}^{\mathrm{A-\psi }}$ represents the
first-order deformation ):
\begin{equation}
\left( S_{1},S_{1}\right) ^{\mathrm{\psi }}=\left( S_{1}^{\mathrm{\psi }%
},S_{1}^{\mathrm{\psi }}\right) +\left( S_{1}^{\mathrm{h-\psi }},S_{1}^{%
\mathrm{h-\psi }}\right) ^{\mathrm{\psi }}+\left( S_{1}^{\mathrm{A-\psi }%
},S_{1}^{\mathrm{A-\psi }}\right) ^{\mathrm{\psi }}.  \label{all2}
\end{equation}%
The notation $\left( S_{1}^{\mathrm{h-\psi }},S_{1}^{\mathrm{h-\psi }%
}\right) ^{\mathrm{\psi }}$ is explained in the above and $\left( S_{1}^{%
\mathrm{A-\psi }},S_{1}^{\mathrm{A-\psi }}\right) ^{\mathrm{\psi }}$ is
discussed in~\cite{SUGRAII}. Let $\Delta ^{\mathrm{\psi }}$ be the
nonintegrated density of $\left( S_{1},S_{1}\right) ^{\mathrm{\psi }}$ and $%
b^{\mathrm{\psi }}$ the nonintegrated density of $S_{2}^{\mathrm{\psi }}$.
Then, (\ref{all1}) takes the local form
\begin{equation}
\Delta ^{\mathrm{\psi }}=-2sb^{\mathrm{\psi }}+\partial ^{\mu }n_{\mu }^{%
\mathrm{\psi }},  \label{all3}
\end{equation}%
with $\mathrm{gh}\left( \Delta ^{\mathrm{\psi }}\right) =1$, $\mathrm{gh}%
\left( b^{\mathrm{\psi }}\right) =0$, $\mathrm{gh}\left( n_{\mu }^{\mathrm{%
\psi }}\right) =1$, for some local currents $n_{\mu }^{\mathrm{\psi }}$. It
is easy to see that $\Delta ^{\mathrm{\psi }}$ decomposes as $\Delta ^{%
\mathrm{\psi }}=\Delta _{0}^{\mathrm{\psi }}+\Delta _{1}^{\mathrm{\psi }%
}+\Delta _{2}^{\mathrm{\psi }}$, with $\mathrm{agh}\left( \Delta _{I}^{%
\mathrm{\psi }}\right) =I$, $I=\overline{0,2}$, where
\begin{equation}
\Delta _{2}^{\mathrm{\psi }}=\bar{\Delta}_{2}^{\mathrm{\psi }},  \label{all4}
\end{equation}%
\begin{equation}
\Delta _{1}^{\mathrm{\psi }}=\bar{\Delta}_{1}^{\mathrm{\psi }}+\tilde{\Delta}%
_{1}^{\mathrm{\psi }},  \label{all5}
\end{equation}%
\begin{eqnarray}
\Delta _{0}^{\mathrm{\psi }} &=&\partial _{\mu }\left( \bar{\tau}_{0}^{\mu }+%
\tilde{\tau}_{0}^{\mu }\right) +\mathrm{i}\left( 180m^{2}-\bar{k}\Lambda
\right) \bar{\xi}\gamma ^{\mu }\psi _{\mu }  \notag \\
&&-\frac{\bar{k}^{2}}{16}\left[ \left( \partial _{\alpha }\bar{\psi}_{\beta
}\right) \gamma ^{\alpha \beta \mu }\psi ^{\nu }\left( \bar{\xi}\gamma
_{(\mu }\psi _{\nu )}-2\sigma _{\mu \nu }\bar{\xi}\gamma ^{\rho }\psi _{\rho
}\right) \right.   \notag \\
&&+\bar{\psi}_{\mu }\gamma ^{\mu \nu \rho }\left( \partial _{\lbrack \nu
}\psi _{\alpha ]}\right) \bar{\xi}\gamma ^{(\alpha }\psi ^{\beta )}\sigma
_{\rho \beta }+\bar{\psi}^{\alpha }\gamma ^{\rho }\psi _{\rho }\partial
^{\beta }\left( \bar{\xi}\gamma _{(\alpha }\psi _{\beta )}\right)   \notag \\
&&\left. +\bar{\psi}^{\alpha }\gamma ^{\rho }\psi ^{\beta }\partial _{\alpha
}\left( \bar{\xi}\gamma _{(\beta }\psi _{\rho )}-2\sigma _{\rho \beta }\bar{%
\xi}\gamma ^{\lambda }\psi _{\lambda }\right) \right]   \notag \\
&&+\mathrm{i}\tilde{k}^{2}\left( \bar{\psi}_{[\mu }\gamma _{\nu \rho }\psi
_{\lambda ]}+\frac{1}{2}\bar{\psi}^{\alpha }\gamma _{\alpha \beta \mu \nu
\rho \lambda }\psi ^{\beta }\right) \partial ^{\mu }\left( \bar{\xi}\gamma
^{\nu \rho }\psi ^{\lambda }\right) .  \label{all6}
\end{eqnarray}%
Related to (\ref{all4}) and (\ref{all5}), we mention that the
expression of $\tilde{\Delta}_{1}^{\mathrm{\psi }}$ can be found
in~\cite{SUGRAII} (they represent the contributions due to the
simultaneous presence of the three-form \textit{and} gravitini), while $%
\bar{\Delta}_{2}^{\mathrm{\psi }}$ and $\bar{\Delta}_{1}^{\mathrm{\psi }}$ are given in (\ref{pfrs4.11}) and (\ref{pfrs4.12}). Since $%
\Delta ^{\mathrm{\psi }}$ has components with the maximum value of
the antighost number equal to two, we can take the nonintegrated
density of the second-order deformation of the solution to the
master equation in the
Rarita-Schwinger sector to stop at antighost number three: $b^{\mathrm{\psi }%
}=\sum\limits_{I=0}^{3}b_{I}^{\mathrm{\psi }}$, $\mathrm{agh}\left( b_{I}^{%
\mathrm{\psi }}\right) =I$, $I=\overline{0,3}$, $n^{\mathrm{\psi }\;\mu
}=\sum\limits_{I=0}^{3}n_{I}^{\mathrm{\psi }\;\mu }$, $\mathrm{agh}\left(
n_{I}^{\mathrm{\psi }\;\mu }\right) =I$, $I=\overline{0,3}$. By projecting (%
\ref{all3}) on various, decreasing values of the antighost numbers, we
obtain that it becomes equivalent to the equations%
\begin{eqnarray}
0 &=&-2\gamma b_{3}^{\mathrm{\psi }}+\partial _{\mu }n_{3}^{\mathrm{\psi \;}%
\mu },  \label{all7a} \\
\Delta _{2}^{\mathrm{\psi }} &=&-2\left( \delta b_{3}^{\mathrm{\psi }%
}+\gamma b_{2}^{\mathrm{\psi }}\right) +\partial _{\mu }n_{2}^{\mathrm{\psi
\;}\mu },  \label{all7b} \\
\Delta _{1}^{\mathrm{\psi }} &=&-2\left( \delta b_{2}^{\mathrm{\psi }%
}+\gamma b_{1}^{\mathrm{\psi }}\right) +\partial _{\mu }n_{1}^{\mathrm{\psi
\;}\mu },  \label{all7c} \\
\Delta _{0}^{\mathrm{\psi }} &=&-2\left( \delta b_{1}^{\mathrm{\psi }%
}+\gamma b_{0}^{\mathrm{\psi }}\right) +\partial _{\mu }n_{0}^{\mathrm{\psi
\;}\mu }.  \label{all7d}
\end{eqnarray}%
Equation (\ref{all7a}) possesses only the trivial solution $b_{3}^{\mathrm{%
\psi }}=0$ (from precisely the same argument as that used in relation with
equation (\ref{pfrs4.16a})). With this result at hand, from (\ref{all4}) and
(\ref{all7b}) we get
\begin{equation}
b_{2}^{\mathrm{\psi }}\equiv \bar{b}_{2}^{\mathrm{\psi }},  \label{all8}
\end{equation}%
where $\bar{b}_{2}^{\mathrm{\psi }}$ is expressed by (\ref{pfrs4.19}). From
the expression of $\Delta _{1}^{\mathrm{\psi }}$ given in (\ref{all5}) we
obtain that (\ref{all7c}) holds if%
\begin{eqnarray}
\chi  &=&\frac{\mathrm{i}\bar{k}^{2}}{8}\psi ^{\ast \mu }\left[ \left(
\partial _{\lbrack \mu }\psi _{\nu ]}\right) \bar{\xi}\gamma ^{\nu }\xi +%
\frac{1}{2}\left( \gamma ^{\nu \rho }\xi \right) \left( \bar{\xi}\gamma
_{\rho }\partial _{\lbrack \mu }\psi _{\nu ]}-\frac{1}{2}\bar{\xi}\gamma
_{\mu }\partial _{\lbrack \nu }\psi _{\rho ]}\right) \right]   \notag \\
&&-\frac{\mathrm{i}\tilde{k}^{2}}{3}\left( \psi _{\lbrack \mu }^{\ast
}\gamma _{\nu \rho \lambda ]}^{\left. {}\right. }\xi -\frac{1}{2}\psi ^{\ast
\sigma }\gamma _{\mu \nu \rho \lambda \sigma }\xi \right) \bar{\xi}\gamma
^{\mu \nu }\partial ^{\lbrack \rho }\psi ^{\lambda ]}  \notag \\
&\equiv &\bar{\chi}+\tilde{\chi},  \label{all9}
\end{eqnarray}%
(with $\bar{\chi}$ as in (\ref{pfrs4.20}) and $\tilde{\chi}$ due to the
presence of the three-form gauge field~\cite{SUGRAII}) decomposes as%
\begin{equation}
\chi =-2\delta \tilde{b}_{2}^{\mathrm{\psi }}+\gamma \rho +\partial _{\mu
}l^{\mu },  \label{all10}
\end{equation}%
where
\begin{eqnarray}
\rho  &=&-\frac{\mathrm{i}\bar{k}^{2}}{16}\psi ^{\ast \mu }\gamma ^{\alpha
\beta }\left( \psi _{\mu }\bar{\xi}\gamma _{\alpha }\psi _{\beta }-\xi
\left( \bar{\psi}_{\mu }\gamma _{\alpha }\psi _{\beta }+\frac{1}{2}\bar{\psi}%
_{\alpha }\gamma _{\mu }\psi _{\beta }\right) \right)   \notag \\
&&+\frac{\mathrm{i}\tilde{k}^{2}}{3}\left( \psi _{\lbrack \mu }^{\ast
}\gamma _{\nu \rho \lambda ]}^{\left. {}\right. }\xi -\frac{1}{2}\psi ^{\ast
\sigma }\gamma _{\mu \nu \rho \lambda \sigma }\xi \right) \bar{\psi}^{\mu
}\gamma ^{\nu \rho }\psi ^{\lambda }-2b_{1}^{\mathrm{\psi }}.  \label{all11}
\end{eqnarray}%
Identities (\ref{fie1}) and (\ref{fie2}) allow us to rewrite (\ref{all9}) in
the form
\begin{eqnarray}
\chi  &=&\frac{\mathrm{i}}{8}\left( \tilde{k}^{2}+\frac{\bar{k}^{2}}{32}%
\right) \left\{ -21\psi ^{\ast \nu }\partial _{\lbrack \mu }\psi _{\nu ]}%
\bar{\xi}\gamma ^{\mu }\xi +\frac{1}{2}\left( -7\psi ^{\ast \alpha }\gamma
_{\mu \nu }^{\;\;\;\beta }\partial _{\lbrack \alpha }\psi _{\beta ]}\right.
\right.   \notag \\
&&\left. +\psi ^{\ast \rho }\gamma _{\mu \nu \rho \alpha \beta }\partial
^{\lbrack \alpha }\psi ^{\beta ]}\right) \bar{\xi}\gamma ^{\mu \nu }\xi -%
\frac{1}{24}\left[ \psi _{\mu }^{\ast }\gamma _{\nu \rho \lambda
}^{\;\;\;\;\varepsilon }\partial _{\lbrack \sigma }\psi _{\varepsilon
]}\right.   \notag \\
&&\left. \left. +\frac{1}{5}\left( \psi ^{\ast \varepsilon }\gamma _{\mu \nu
\rho \lambda \sigma \alpha \beta }-\psi _{\alpha }^{\ast }\gamma _{\mu \nu
\rho \lambda \sigma \beta }\right) \partial ^{\lbrack \alpha }\psi ^{\beta ]}%
\right] \bar{\xi}\gamma ^{\mu \nu \rho \lambda \sigma }\xi \right\} +\delta
\Omega ,  \label{all12}
\end{eqnarray}%
where we made the notation
\begin{eqnarray}
\Omega  &=&\left[ \frac{1}{8}\left( 3\tilde{k}^{2}-\frac{79\bar{k}^{2}}{%
9\cdot 2^{7}}\right) \psi _{\nu }^{\ast }\gamma ^{\mu }\bar{\psi}^{\ast \nu
}-\frac{1}{16}\left( \tilde{k}^{2}-\frac{11\bar{k}^{2}}{9\cdot 2^{6}}\right)
\psi _{\nu }^{\ast }\gamma ^{\mu \nu \rho }\bar{\psi}_{\rho }^{\ast }\right.
\notag \\
&&\left. -\frac{1}{6}\left( \tilde{k}^{2}-\frac{7\bar{k}^{2}}{3\cdot 2^{8}}%
\right) \psi _{\nu }^{\ast }\gamma ^{\nu }\bar{\psi}^{\ast \mu }\right] \bar{%
\xi}\gamma _{\mu }\xi +\frac{1}{8}\left[ \frac{1}{2}\left( 13\tilde{k}^{2}-%
\frac{25\bar{k}^{2}}{9\cdot 2^{6}}\right) \psi _{\mu }^{\ast }\bar{\psi}%
_{\nu }^{\ast }\right.   \notag \\
&&+\frac{1}{4}\left( \tilde{k}^{2}+\frac{11\bar{k}^{2}}{9\cdot 2^{6}}\right)
\psi ^{\ast \rho }\gamma _{\mu \nu \rho \lambda }\bar{\psi}^{\ast \lambda }-%
\frac{4}{3}\left( \tilde{k}^{2}+\frac{7\bar{k}^{2}}{3\cdot 2^{8}}\right)
\psi _{\mu }^{\ast }\gamma _{\nu \rho }\bar{\psi}^{\ast \rho }  \notag \\
&&\left. -\frac{1}{2}\left( 3\tilde{k}^{2}+\frac{79\bar{k}^{2}}{9\cdot 2^{7}}%
\right) \psi ^{\ast \rho }\gamma _{\mu \nu }\bar{\psi}_{\rho }^{\ast }\right]
\bar{\xi}\gamma ^{\mu \nu }\xi   \notag \\
&&+\frac{1}{2^{3}\cdot 3^{3}}\left[ \frac{7}{2}\left( \tilde{k}^{2}+\frac{%
\bar{k}^{2}}{2^{6}}\right) \psi _{\mu }^{\ast }\gamma _{\nu \rho \lambda }%
\bar{\psi}_{\sigma }^{\ast }+\frac{1}{8}\left( \tilde{k}^{2}+\frac{5\bar{k}%
^{2}}{2^{7}}\right) \psi ^{\ast \varepsilon }\gamma _{\mu \nu \rho \lambda
\sigma }\bar{\psi}_{\varepsilon }^{\ast }\right.   \notag \\
&&\left. -\frac{7}{8}\left( \tilde{k}^{2}+\frac{\bar{k}^{2}}{2^{6}}\right)
\psi _{\sigma }^{\ast }\gamma _{\mu \nu \rho \lambda \varepsilon }\bar{\psi}%
^{\ast \varepsilon }-\frac{1}{10}\left( \tilde{k}^{2}+\frac{11\bar{k}^{2}}{%
2^{9}}\right) \psi ^{\ast \varepsilon }\gamma _{\mu \nu \rho \lambda \sigma
\varepsilon \eta }\bar{\psi}^{\ast \eta }\right] \times   \notag \\
&&\times \bar{\xi}\gamma ^{\mu \nu \rho \lambda \sigma }\xi .  \label{al13}
\end{eqnarray}%
Reprising the procedure employed in the previous section between formulas (%
\ref{pfrs4.21}) and (\ref{pfrs4.22}), we deduce that
\begin{eqnarray}
\tilde{k}^{2}+\frac{\bar{k}^{2}}{32} &=&0,  \label{all14a} \\
\tilde{b}_{2}^{\mathrm{\psi }} &=&-\frac{1}{2}\Omega .  \label{all14b}
\end{eqnarray}%
Consequently, (\ref{all10}) takes the simple form
\begin{equation}
\gamma \rho +\partial _{\mu }l^{\mu }=0,  \label{all24a}
\end{equation}%
which, since $\mathrm{agh}\left( \rho \right) =1>0$, can be replaced with%
\begin{equation}
\gamma \rho =0.  \label{gamarho=0}
\end{equation}%
Inserting (\ref{all14b}) in (\ref{all7c}) we obtain the component of
antighost number one from the second-order deformation of the solution to
the master equation in the Rarita-Schwinger sector as
\begin{eqnarray*}
b_{1}^{\mathrm{\psi }} &=&-\frac{\mathrm{i}\bar{k}^{2}}{32}\psi ^{\ast \mu
}\gamma ^{\alpha \beta }\left( \psi _{\mu }\bar{\xi}\gamma _{\alpha }\psi
_{\beta }-\xi \left( \bar{\psi}_{\mu }\gamma _{\alpha }\psi _{\beta }+\frac{1%
}{2}\bar{\psi}_{\alpha }\gamma _{\mu }\psi _{\beta }\right) \right)  \\
&&+\frac{\mathrm{i}\tilde{k}^{2}}{6}\left( \psi _{\lbrack \mu }^{\ast
}\gamma _{\nu \rho \lambda ]}^{\left. {}\right. }\xi -\frac{1}{2}\psi ^{\ast
\sigma }\gamma _{\mu \nu \rho \lambda \sigma }\xi \right) \bar{\psi}^{\mu
}\gamma ^{\nu \rho }\psi ^{\lambda }.
\end{eqnarray*}%
Expressing $\tilde{k}^{2}$ in terms of $\bar{k}^{2}$ with the help of (\ref%
{all14a}) we find that the pieces of antighost number two and one from the
second-order deformation, responsible for the self-interactions among
gravitini, become parameterized by a single constant
\begin{eqnarray}
b_{2}^{\mathrm{\psi }} &=&-\frac{\mathrm{i}\bar{k}^{2}}{32}\xi ^{\ast
}\gamma ^{\mu \nu }\xi \bar{\xi}\gamma _{\mu }\psi _{\nu }-\frac{\bar{k}^{2}%
}{9\cdot 2^{10}}\left( -\frac{11\cdot 17}{2}\psi _{\nu }^{\ast }\gamma ^{\mu
}\bar{\psi}^{\ast \nu }\right.   \notag \\
&&\left. +\frac{29}{2}\psi _{\nu }^{\ast }\gamma ^{\mu \nu \rho }\bar{\psi}%
_{\rho }^{\ast }+31\psi _{\nu }^{\ast }\gamma ^{\nu }\bar{\psi}^{\ast \mu
}\right) \bar{\xi}\gamma _{\mu }\xi   \notag \\
&&-\frac{\bar{k}^{2}}{9\cdot 2^{10}}\left( -\frac{7\cdot 37}{2}\psi _{\mu
}^{\ast }\bar{\psi}_{\nu }^{\ast }-\frac{7}{4}\psi ^{\ast \rho }\gamma _{\mu
\nu \rho \lambda }\bar{\psi}^{\ast \lambda }+17\psi _{\mu }^{\ast }\gamma
_{\nu \rho }\bar{\psi}^{\ast \rho }\right.   \notag \\
&&\left. +\frac{29}{4}\psi ^{\ast \rho }\gamma _{\mu \nu }\bar{\psi}_{\rho
}^{\ast }\right) \bar{\xi}\gamma ^{\mu \nu }\xi   \notag \\
&&-\frac{\bar{k}^{2}}{27\cdot 2^{11}}\left( -7\psi _{\mu }^{\ast }\gamma
_{\nu \rho \lambda }\bar{\psi}_{\sigma }^{\ast }+\frac{1}{8}\psi ^{\ast
\varepsilon }\gamma _{\mu \nu \rho \lambda \sigma }\bar{\psi}_{\varepsilon
}^{\ast }\right.   \notag \\
&&\left. +\frac{7}{4}\psi _{\sigma }^{\ast }\gamma _{\mu \nu \rho \lambda
\varepsilon }\bar{\psi}^{\ast \varepsilon }+\frac{1}{8}\psi ^{\ast
\varepsilon }\gamma _{\mu \nu \rho \lambda \sigma \varepsilon \eta }\bar{\psi%
}^{\ast \eta }\right) \bar{\xi}\gamma ^{\mu \nu \rho \lambda \sigma }\xi ,
\label{al15}
\end{eqnarray}%
\begin{eqnarray}
b_{1}^{\mathrm{\psi }} &=&-\frac{\mathrm{i}\bar{k}^{2}}{32}\left\{ \psi
^{\ast \mu }\gamma ^{\alpha \beta }\left[ \psi _{\mu }\bar{\xi}\gamma
_{\alpha }\psi _{\beta }-\xi \left( \bar{\psi}_{\mu }\gamma _{\alpha }\psi
_{\beta }+\frac{1}{2}\bar{\psi}_{\alpha }\gamma _{\mu }\psi _{\beta }\right) %
\right] \right.   \notag \\
&&\left. +\frac{1}{6}\left( \psi _{\lbrack \mu }^{\ast }\gamma _{\nu \rho
\lambda ]}^{\left. {}\right. }\xi -\frac{1}{2}\psi ^{\ast \sigma }\gamma
_{\mu \nu \rho \lambda \sigma }\xi \right) \bar{\psi}^{\mu }\gamma ^{\nu
\rho }\psi ^{\lambda }\right\} .  \label{all16}
\end{eqnarray}%
Relation (\ref{all14a}) and (\ref{all6}) lead, by direct computation, to%
\begin{eqnarray}
\Delta _{0}^{\mathrm{\psi }} &=&\partial _{\mu }\left( \bar{\tau}_{0}^{\mu }+%
\tilde{\tau}_{0}^{\mu }\right) +\mathrm{i}\left( 180m^{2}-\bar{k}\Lambda
\right) \bar{\xi}\gamma ^{\mu }\psi _{\mu }-2\delta b_{1}^{\mathrm{\psi }}
\notag \\
&&+\gamma \left\{ \frac{\bar{k}^{2}}{16}\left[ \bar{\psi}^{\alpha }\gamma
^{\mu }\psi _{\mu }\bar{\psi}_{\alpha }\gamma ^{\nu }\psi _{\nu }-\frac{1}{4}%
\bar{\psi}_{\alpha }\gamma _{\rho }\psi _{\beta }\left( \bar{\psi}^{\alpha
}\gamma ^{\rho }\psi ^{\beta }\right. \right. \right.   \notag \\
&&\left. +2\bar{\psi}^{\alpha }\gamma ^{\beta }\psi ^{\rho }+\bar{\psi}_{\mu
}\gamma ^{\mu \nu \rho \alpha \beta }\psi _{\nu }\right)   \notag \\
&&-\frac{1}{8}\bar{\psi}_{\mu }\gamma _{\nu \rho }\psi _{\lambda }\left(
\bar{\psi}^{[\mu }\gamma ^{\nu \rho }\psi ^{\lambda ]}+\bar{\psi}_{\alpha
}\gamma ^{\alpha \beta \mu \nu \rho \lambda }\psi _{\beta }\right)   \notag
\\
&&-\frac{1}{60}\left( \bar{\psi}_{\mu }\gamma ^{\mu \nu \rho \lambda
\alpha }\psi ^{\sigma }\bar{\psi}_{\alpha }\gamma _{\nu \rho \lambda
\sigma \beta }\psi ^{\beta }+\frac{1}{4}\bar{\psi}^{\mu }\gamma
_{\mu \nu \rho \lambda \sigma }\psi ^{\nu }\bar{\psi}_{\alpha
}\gamma ^{\alpha \beta \rho \lambda
\sigma }\psi _{\beta }\right.   \notag \\
&&\left. \left. \left. +\frac{1}{4}\bar{\psi}^{\mu }\gamma ^{\alpha
\beta \rho \lambda \sigma }\psi ^{\nu }\bar{\psi}_{\alpha }\gamma
_{\mu \nu \rho \lambda \sigma }\psi _{\beta }\right) \right]
\right\} ,  \label{al17}
\end{eqnarray}%
where $b_{1}^{\mathrm{\psi }}$ reads as in (\ref{all16}). Comparing the
right-hand sides of (\ref{all7d}) and (\ref{al17}) we finally identify the
second-order Lagrangian in the Rarita-Schwinger sector, which indeed
includes quartic gravitini vertices%
\begin{eqnarray}
b_{0}^{\mathrm{\psi }} &=&-\frac{\bar{k}^{2}}{32}\left[ \bar{\psi}^{\alpha
}\gamma ^{\mu }\psi _{\mu }\bar{\psi}_{\alpha }\gamma ^{\nu }\psi _{\nu }-%
\frac{1}{4}\bar{\psi}_{\alpha }\gamma _{\rho }\psi _{\beta }\left( \bar{\psi}%
^{\alpha }\gamma ^{\rho }\psi ^{\beta }\right. \right.   \notag \\
&&\left. +2\bar{\psi}^{\alpha }\gamma ^{\beta }\psi ^{\rho }+\bar{\psi}_{\mu
}\gamma ^{\mu \nu \rho \alpha \beta }\psi _{\nu }\right)   \notag \\
&&-\frac{1}{8}\bar{\psi}_{\mu }\gamma _{\nu \rho }\psi _{\lambda }\left(
\bar{\psi}^{[\mu }\gamma ^{\nu \rho }\psi ^{\lambda ]}+\bar{\psi}_{\alpha
}\gamma ^{\alpha \beta \mu \nu \rho \lambda }\psi _{\beta }\right)   \notag
\\
&&-\frac{1}{60}\left( \bar{\psi}_{\mu }\gamma ^{\mu \nu \rho \lambda
\alpha }\psi ^{\sigma }\bar{\psi}_{\alpha }\gamma _{\nu \rho \lambda
\sigma \beta }\psi ^{\beta }+\frac{1}{4}\bar{\psi}^{\mu }\gamma
_{\mu \nu \rho \lambda \sigma }\psi ^{\nu }\bar{\psi}_{\alpha
}\gamma ^{\alpha \beta \rho \lambda
\sigma }\psi _{\beta }\right.   \notag \\
&&\left. \left. +\frac{1}{4}\bar{\psi}^{\mu }\gamma ^{\alpha \beta
\rho \lambda \sigma }\psi ^{\nu }\bar{\psi}_{\alpha }\gamma _{\mu
\nu \rho \lambda \sigma }\psi _{\beta }\right) \right] ,
\label{al18}
\end{eqnarray}%
and also deduce the identity
\begin{equation}
180m^{2}-\bar{k}\Lambda =0.  \label{al19}
\end{equation}%
Thus, we have shown that it is precisely the presence of \textit{all three}
types of fields which induces the appearance of quartic self-interactions
among gravitini of the type (\ref{al18}). The main argument is quite clear
now: if one considers the spin-$3/2$ field coupled to either a spin-two
field or a three-form, then (\ref{all14a}) reduces to either $\bar{k}^{2}=0$
or respectively to $\tilde{k}^{2}=0$, which gives $\bar{b}_{0}^{\mathrm{\psi
}}=0$ or respectively $\tilde{b}_{0}^{\mathrm{\psi }}=0$. Another
interesting remark is that equation (\ref{al19}) apparently allows the
presence of cosmological and gravitini `mass' constants, which are known to
be forbidden in $D=11$, $N=1$ SUGRA. We will see in~\cite{SUGRAIV} that
equation (\ref{al19}) actually sets zero both constants, $\Lambda =0=m$.

\section{Comparison with the case $D=4$}

It is useful to make a short comparison between the cases $D=11$ and $D=4$
since it is known that the same free theory (describing a graviton and a
massless $3/2$-field) allows for self-interactions among gravitini in $D=4$
and, meanwhile, forces an algebraic relation between the gravitini `mass'
constant and the cosmological one. In four dimensions the first-order
deformation of the solution to the master equation takes a form similar to $%
D=11$, excepting the `mass' terms, which are written as $a_{\mathrm{D=4}}^{%
\mathrm{\psi }}=m\left( \mathrm{i}\psi _{\mu }^{\ast }\gamma ^{\mu }\xi -%
\bar{\psi}_{\mu }\gamma ^{\mu \nu }\psi _{\nu }\right) $. Similarly to the
case $D=11$, we will denote by $\Delta _{\mathrm{D=4}}^{\mathrm{\psi }}$ and
$b_{\mathrm{D=4}}^{\mathrm{\psi }}$ the nonintegrated densities of $\left(
S_{1},S_{1}\right) ^{\mathrm{\psi }}$ and of $S_{2}^{\mathrm{\psi }}$
respectively. The consistency of the first-order deformation in the
Rarita-Schwinger sector is equivalent to the equation $\Delta _{\mathrm{D=4}%
}^{\mathrm{\psi }}=-2sb_{\mathrm{D=4}}^{\mathrm{\psi }}+\partial _{\mu }n_{%
\mathrm{D=4}}^{\mu }$, where $\Delta _{\mathrm{D=4}}^{\mathrm{\psi }}=\Delta
_{\mathrm{D=4},2}^{\mathrm{\psi }}+\Delta _{\mathrm{D=4},1}^{\mathrm{\psi }%
}+\Delta _{\mathrm{D=4},0}^{\mathrm{\psi }}$, with $\Delta _{\mathrm{D=4}%
,2}^{\mathrm{\psi }}$ and $\Delta _{\mathrm{D=4},1}^{\mathrm{\psi }}$ having
exactly the expressions (\ref{pfrs4.11}) and (\ref{pfrs4.12}), but in $D=4$,
and $\Delta _{\mathrm{D=4},0}^{\mathrm{\psi }}$ being given by
\begin{eqnarray}
&&\Delta _{\mathrm{D=4},0}^{\mathrm{\psi }}=\mathrm{i}\left( 12m^{2}-\bar{k}%
\Lambda \right) \bar{\xi}\gamma ^{\mu }\psi _{\mu }  \notag \\
&&-\frac{\bar{k}^{2}}{16}\left[ \left( \partial _{\alpha }\bar{\psi}_{\beta
}\right) \gamma ^{\alpha \beta \mu }\psi ^{\nu }\left( \bar{\xi}\gamma
_{(\mu }\psi _{\nu )}-2\sigma _{\mu \nu }\bar{\xi}\gamma ^{\rho }\psi _{\rho
}\right) \right.   \notag \\
&&+\bar{\psi}_{\mu }\gamma ^{\mu \nu \rho }\left( \partial _{\lbrack \nu
}\psi _{\alpha ]}\right) \bar{\xi}\gamma ^{(\alpha }\psi ^{\beta )}\sigma
_{\rho \beta }+\bar{\psi}^{\alpha }\gamma ^{\rho }\psi _{\rho }\partial
^{\beta }\left( \bar{\xi}\gamma _{(\alpha }\psi _{\beta )}\right)   \notag \\
&&\left. +\bar{\psi}^{\alpha }\gamma ^{\rho }\psi ^{\beta }\partial _{\alpha
}\left( \bar{\xi}\gamma _{(\beta }\psi _{\rho )}-2\sigma _{\rho \beta }\bar{%
\xi}\gamma ^{\lambda }\psi _{\lambda }\right) \right] .  \label{pr3}
\end{eqnarray}%
The key point in $D=4$ is that the analogue of $\bar{\chi}$ expressed by (%
\ref{pfrs4.20}) can be rewritten as
\begin{eqnarray}
&&\chi _{\mathrm{D=4}}=\partial _{\mu }\theta ^{\mu }+\delta \left[ -\frac{1%
}{2^{7}}\bar{\xi}\gamma _{\mu }\xi \left( \psi _{\nu }^{\ast }\gamma ^{\mu }%
\bar{\psi}^{\ast \nu }+\frac{3}{2}\psi _{\nu }^{\ast }\gamma ^{\mu \nu \rho }%
\bar{\psi}_{\rho }^{\ast }\right) \right.   \notag \\
&&\left. +\frac{1}{2^{7}}\bar{\xi}\gamma _{\mu \nu }\xi \left( \psi _{\rho
}^{\ast }\gamma ^{\mu \nu \rho \lambda }\bar{\psi}_{\lambda }^{\ast }-\frac{1%
}{2}\psi ^{\ast \rho }\gamma ^{\mu \nu }\bar{\psi}_{\rho }^{\ast }-2\psi
^{\ast \mu }\bar{\psi}^{\ast \nu }\right) \right] .  \label{pr4}
\end{eqnarray}%
With these result at hand we find in $D=4$ that $b_{\mathrm{D=4},3}^{\mathrm{%
\psi }}=0$ and
\begin{eqnarray}
b_{\mathrm{D=4},2}^{\mathrm{\psi }} &=&-\frac{\mathrm{i}\bar{k}^{2}}{32}\xi
^{\ast }\gamma ^{\mu \nu }\xi \bar{\xi}\gamma _{\mu }\psi _{\nu }+\frac{1}{%
2^{8}}\bar{\xi}\gamma _{\mu }\xi \left( \psi _{\nu }^{\ast }\gamma ^{\mu }%
\bar{\psi}^{\ast \nu }+\frac{3}{2}\psi _{\nu }^{\ast }\gamma ^{\mu \nu \rho }%
\bar{\psi}_{\rho }^{\ast }\right)   \notag \\
&&-\frac{1}{2^{8}}\bar{\xi}\gamma _{\mu \nu }\xi \left( \psi _{\rho }^{\ast
}\gamma ^{\mu \nu \rho \lambda }\bar{\psi}_{\lambda }^{\ast }-\frac{1}{2}%
\psi ^{\ast \rho }\gamma ^{\mu \nu }\bar{\psi}_{\rho }^{\ast }-2\psi ^{\ast
\mu }\bar{\psi}^{\ast \nu }\right) ,  \label{pr5}
\end{eqnarray}%
\begin{equation}
b_{\mathrm{D=4},1}^{\mathrm{\psi }}=-\frac{\mathrm{i}\bar{k}^{2}}{32}\left[
\psi ^{\ast \mu }\gamma ^{\alpha \beta }\psi _{\mu }\bar{\xi}\gamma _{\alpha
}\psi _{\beta }-\psi ^{\ast \mu }\gamma ^{\alpha \beta }\xi \left( \bar{\psi}%
_{\mu }\gamma _{\alpha }\psi _{\beta }+\frac{1}{2}\bar{\psi}_{\alpha }\gamma
_{\mu }\psi _{\beta }\right) \right] ,  \label{pr6}
\end{equation}%
while (\ref{pr3}) can be put in the form
\begin{eqnarray}
&&\Delta _{\mathrm{D=4},0}^{\mathrm{\psi }}=\mathrm{i}\left( 12m^{2}-\bar{k}%
\Lambda \right) \bar{\xi}\gamma ^{\mu }\psi _{\mu }  \notag \\
&&+\delta \left\{ \frac{\mathrm{i}\bar{k}^{2}}{16}\left[ \psi ^{\ast \mu
}\gamma ^{\alpha \beta }\psi _{\mu }\bar{\xi}\gamma _{\alpha }\psi _{\beta
}-\psi ^{\ast \mu }\gamma ^{\alpha \beta }\xi \left( \bar{\psi}_{\mu }\gamma
_{\alpha }\psi _{\beta }+\frac{1}{2}\bar{\psi}_{\alpha }\gamma _{\mu }\psi
_{\beta }\right) \right] \right\}   \notag \\
&&+\gamma \left[ \frac{\bar{k}^{2}}{16}\left( \bar{\psi}^{\alpha }\gamma
^{\mu }\psi _{\mu }\bar{\psi}_{\alpha }\gamma ^{\nu }\psi _{\nu }-\frac{1}{4}%
\bar{\psi}^{\alpha }\gamma ^{\mu }\psi ^{\beta }\left( \bar{\psi}_{\alpha
}\gamma _{\mu }\psi _{\beta }+2\bar{\psi}_{\alpha }\gamma _{\beta }\psi
_{\mu }\right) \right) \right] .  \label{pr7}
\end{eqnarray}%
Relation (\ref{pr7}) provides the piece of antighost number zero from the
second-order deformation of the solution to the master equation in the
Rarita-Schwinger field
\begin{equation}
b_{\mathrm{D=4},\;0}^{\mathrm{\psi }}=-\frac{\bar{k}^{2}}{32}\left( \bar{\psi%
}^{\alpha }\gamma ^{\mu }\psi _{\mu }\bar{\psi}_{\alpha }\gamma ^{\nu }\psi
_{\nu }-\frac{1}{4}\bar{\psi}^{\alpha }\gamma ^{\mu }\psi ^{\beta }\left(
\bar{\psi}_{\alpha }\gamma _{\mu }\psi _{\beta }+2\bar{\psi}_{\alpha }\gamma
_{\beta }\psi _{\mu }\right) \right)   \label{pr8}
\end{equation}%
and also enforces the condition $12m^{2}-\bar{k}\Lambda =0$, which expresses
the well-known relation between the coupling constants $m$ and $\bar{k}$ and
the cosmological constant $\Lambda $. Finally, we remark that it is the same
object, namely (\ref{pfrs4.20}), which in $D=4$ does satisfy an equation of
the type (\ref{mn3}) (see (\ref{pr4})) and thus ensures the existence of
interactions, but in $D=11$ cannot satisfy such an equation and consequently
forbids the presence of interactions.

\section{Conclusion}

To conclude with, in this paper we have investigated the
eleven-dimensional couplings between a massless spin-two field
(described in the free limit by a Pauli-Fierz action) and a massless
Rarita-Schwinger spinor using the powerful setting based on local
BRST cohomology. Under the hypotheses of
locality, smoothness of the interactions in the coupling constant, Poincar%
\'{e} invariance, Lorentz covariance, and the preservation of the number of
derivatives on each field, we have shown that in $D=11$ there are no
consistent cross-interactions among the graviton and the massless gravitino
and also no self-interactions in the Rarita-Schwinger sector, unlike the
case $D=4$, where such cross-interactions exist.

\section*{Acknowledgments}

The authors wish to thank Constantin Bizdadea and Odile Saliu for useful
discussions and comments. This work is partially supported by the European
Commission FP6 program MRTN-CT-2004-005104 and by the grant AT24/2005 with
the Romanian National Council for Academic Scientific Research
(C.N.C.S.I.S.) and the Romanian Ministry of Education and Research (M.E.C.).

\end{document}